\documentclass[preprint,amsmath,amssymb,aps,showkeys,showpacs]{revtex4}
\usepackage[english]{babel}
\usepackage{graphicx}
\usepackage{graphics}
\usepackage{amsmath}
\usepackage{dcolumn}
\usepackage{amssymb}
\usepackage{bm}

\begin{document}
\title{Quantum effects on an atom with a magnetic quadrupole moment in a region with a time-dependent magnetic field}
\author{I. C. Fonseca}
\affiliation{Departamento de F\'isica, Universidade Federal da Para\'iba, Caixa Postal 5008, 58051-900, Jo\~ao Pessoa-PB, Brazil.}

\author{K. Bakke}
\email{kbakke@fisica.ufpb.br}
\affiliation{Departamento de F\'isica, Universidade Federal da Para\'iba, Caixa Postal 5008, 58051-900, Jo\~ao Pessoa-PB, Brazil.}

\begin{abstract}
The quantum description of an atom with a magnetic quadrupole moment in the presence of a time-dependent magnetic field is analysed. It is shown that the time-dependent magnetic field induces an electric field that interacts with the magnetic quadrupole moment of the atom and gives rise to a Landau-type quantization. It is also shown that a time-independent Schr\"odinger equation can be obtained, i.e., without existing the interaction between the magnetic quadrupole moment of the atom and the time-dependent magnetic field, therefore, the Schr\"odinger equation can be solved exactly. It is also analysed this system subject to scalar potentials.
\end{abstract}


\maketitle

\section{Introduction}

Neutral particle systems have been widely explored in recent decades due to the interactions between magnetic and electric dipole moments and the electromagnetic field. Well-known examples of the interaction of the permanent magnetic dipole moment with an electric field are the Aharonov-Casher geometric quantum phase \cite{ac} and the analogue of the Landau quantization \cite{er}. Atoms with an induced electric dipole moment that interacts with electric and magnetic fields have also shown quantum effects associated with geometric quantum phases \cite{whw} and the Landau-type quantization \cite{lin3}. Besides, by dealing with neutral particles with a permanent electric dipole moment that interacts with a magnetic field, it has been shown in Refs. \cite{hmw,hmw2} that geometric quantum phases can be obtained. In the context of quantum field theory, neutral particle systems have been used to study effects of the violation of the Lorentz symmetry at low energy systems \cite{book}, and thus obtain quantum effects such as analogues of the quantum Hall effect \cite{lin2} and the Rashba coupling \cite{rash}. Other neutral particle systems that have been explored in the literature take into account the electric and magnetic quadrupole moments. The quadrupole moments of atoms and molecules have been the focus of studies of single crystals \cite{quad}, molecules \cite{quad-1,quad-2,quad5,quad15,quad17}, atoms \cite{quad6,prlquad}, chiral states \cite{quad18}, geometric quantum phases \cite{chen,fb2} and noncommutative quantum mechanics \cite{nonc}. In particular, quantum particles with a magnetic quadrupole moment have attracted interests in atomic systems \cite{magq5,magq6}, molecules \cite{magq7,magq9}, chiral anomaly \cite{magq}, Landau-type quantization \cite{fb2,fb4} and the analogue of the quantum Hall effect \cite{fb3}. Other interesting studies have been made in Refs. \cite{magq1,magq2,pra,prc,magq4}.

In this work, we consider the single particle approximation used in Refs. \cite{prc,pra,fb}. Then, we deal with a system that consists in a moving atom with a magnetic quadrupole moment that interacts with external fields. The aim of this work is to analyse the behaviour of an atom with a magnetic quadrupole moment in the presence of a time-dependent magnetic field. We show that an induced electric field interacts with the magnetic quadrupole moment and gives rise to a Landau-type quantization. Besides, this analogue of the Landau quantization is achieved without the interaction between the magnetic quadrupole moment of the atom and the time-dependent magnetic field. We also analyse the influence of a hard-wall confining potential, a scalar potential proportional to the inverse of the radial distance and a linear scalar potential on this system, and thus, fill a lack in studies of neutral particle systems that possess multipole moments.

The structure of this work is as follows: in section II, we discuss the analogue of the Landau quantization for an atom with a magnetic quadrupole moment in the presence of a time-dependent magnetic field; in section III, we discuss the case of this neutral particle in the presence of a time-dependent magnetic field subject to a hard-wall confining potential; in section IV, we analyse the neutral particle in the presence of a time-dependent magnetic field subject to a scalar potential proportional to the inverse of the radial distance; in section V, we investigate the behaviour of the neutral particle in the presence of a time-dependent magnetic field subject to a linear scalar potential; in section VI, we present our conclusions.

\section{Landau-type quantization}

In this section, we start by making a brief review of the quantum description of a neutral particle (atom or molecule) with a magnetic quadrupole moment that interacts with magnetic and electric field proposed in Ref. \cite{fb}. In the following, we show a particular case where an atom with a magnetic quadrupole moment moves in a region in which there exists a time-dependent magnetic field, but the interaction between the magnetic quadrupole moment and this time-dependent magnetic field gives rise to an analogue of the Landau quantization, that is, we can obtain and solve a time-independent Schr\"odinger equation exactly. Based on Refs. \cite{pra,prc}, an atom or a molecule with a magnetic quadrupole moment can be seen as a scalar particle. If the particle moves with velocity $v\ll c$, the quantum description of a neutral particle with a magnetic quadrupole moment can be described by the Schr\"odinger equation (with SI units)
\begin{eqnarray}
i\hbar\frac{\partial\psi}{\partial t}=\frac{1}{2m}\left[\hat{p}-\frac{1}{c^{2}}(\vec{M}\times\vec{E})\right]^2\,\psi-\vec{M}\cdot\vec{B}\,\psi,
\label{1.1}
\end{eqnarray} 
where the vector $\vec{M}$ has the components determined by $M_{i}=\sum_{j}M_{ij}\,\partial_{j}$, where $M_{ij}$ is the magnetic quadrupole moment tensor (symmetric and traceless tensor) and the fields $\vec{E}$ and $\vec{B}$ are the electric and magnetic fields in the laboratory frame, respectively \cite{fb,fb5,griff}.

On the other hand, the Landau quantization arises from the interaction between an electrically charged particle with a uniform magnetic field \cite{landau}. A Landau-type quantization is achieved for an atom with a magnetic quadrupole moment when there is a uniform effective magnetic field perpendicular to the plane of motion of the particle given by
\begin{eqnarray}
\vec{B}_{\mathrm{eff}}=\vec{\nabla}\times\left[\vec{M}\times\vec{E}\right],
\label{5.1}
\end{eqnarray}
 where the vector $\vec{E}$ is the electric field in the laboratory frame and satisfies the electrostatic conditions \cite{fb2,fb3}.

Now, since the tensor $M_{ij}$ is symmetric and traceless, let us consider the magnetic quadrupole moment tensor to be defined by the following components:
\begin{eqnarray}
M_{\rho\rho}=M_{\varphi\varphi}=M;\,\,\,\,\,M_{zz}=-2M,
\label{5.1a}
\end{eqnarray}
where $M$ is a constant $\left(M>0\right)$ and all other components of $M_{ij}$ are null. Further, let us consider an electric field given by
\begin{eqnarray}
\vec{E}=\frac{b\,\rho^{2}}{2}\,\hat{z},
\label{5.2}
\end{eqnarray}
where $b$ is a constant, $\rho=\sqrt{x^{2}+y^{2}}$ is the radial coordinate and $\hat{z}$ is a unit vector in the $z$-direction. In this way, we have an effective vector potential given by $\vec{A}_{\mathrm{eff}}=\vec{M}\times\vec{E}=-b\,M\,\rho\,\hat{\varphi}$ ($\hat{\varphi}$ is a unit vector in the azimuthal direction) and, consequently, the effective magnetic field (\ref{5.1}) is uniform in the $z$-direction, that is, it is perpendicular to the plane of motion of the quantum particle. However, the electric field given in Eq. (\ref{5.2}) does not satisfy the electrostatic conditions since $\vec{\nabla}\times\vec{E}\neq0$. Therefore, there exists a time-dependent magnetic field $\vec{B}=B\left(t,\rho\right)\,\hat{\varphi}=b\,\rho\,t\,\hat{\varphi}$ which breaks one of the conditions of achieving the Landau quantization for a magnetic quadrupole moment. Note that the direction of the electric field (\ref{5.2}) differs from that used in Ref. \cite{fb2}, where the electric field is in the radial direction and it is produced by a non-uniform distribution of electric charges inside a non-conductor cylinder. Besides, the structure of the magnetic quadrupole tensor considered in Ref. \cite{fb2} differs from that of Eq. (\ref{5.1a}). This means that the interaction of the induced electric field (\ref{5.2}) with an atom that possesses a non-diagonal magnetic quadrupole tensor as considered in Ref. \cite{fb2} would be null, and thus the Landau-type quantization would not be achieved. This difference between the present proposal and that of Ref. \cite{fb2} is subtle and it shows us that the choice of the electric field depends on the structure of the magnetic quadrupole moment of the atom in order that the Landau quantization can be obtained. 

On the other hand, the time-dependent magnetic field $\vec{B}=b\,\rho\,t\,\hat{\varphi}$ does not interact with the magnetic quadrupole moment defined by the components given in Eq. (\ref{5.1a}), that is, we have that $\vec{M}\cdot\vec{B}=0$. Therefore, the last term of Eq. (\ref{1.1}) vanishes and the Schr\"odinger equation (\ref{1.1}) becomes (by working with $\hbar=c=1$ from now on):
\begin{eqnarray}
i\frac{\partial\psi}{\partial t}=-\frac{1}{2m}\left[\frac{\partial^{2}}{\partial\rho^{2}}+\frac{1}{\rho}\,\frac{\partial}{\partial\rho}+\frac{1}{\rho^{2}}\,\frac{\partial^{2}}{\partial\varphi^{2}}+\frac{\partial^{2}}{\partial z}\right]\psi+i\frac{M\,b}{m}\,\frac{\partial\psi}{\partial\varphi}+\frac{M^{2}\,b^{2}}{2m}\,\rho^{2}\,\psi.
\label{5.3}
\end{eqnarray}
A solution to Eq. (\ref{5.3}) can be written through the ansatz: $\psi=e^{-i\mathcal{E}t}\,e^{i\,l\,\varphi}\,e^{ikz}\,R\left(\rho\right)$, where $l=0,\pm1,\pm2,\ldots$ and $k$ is a constant. By substituting this solution into Eq. (\ref{5.3}), we obtain a time-independent equation given by:
\begin{eqnarray}
\left[2m\mathcal{E}-k^{2}\right]R=-R''-\frac{1}{\rho}\,R'+\frac{l^{2}}{\rho^{2}}\,R-2\,M\,b\,l\,R+M^{2}\,b^{2}\,\rho^{2}\,R.
\label{5.4}
\end{eqnarray}

Henceforth, we reduce the system to a planar system by taking $k=0$. In what follows, we make a simple change of variables in Eq. (\ref{5.4}) given by: $\xi=M\,b\,\rho^{2}$. Thus, we have
\begin{eqnarray}
\xi\,R''+R'-\frac{l^{2}}{4\xi}\,R-\frac{\xi}{4}\,R+\mu\,R=0,
\label{5.5}
\end{eqnarray} 
where we have defined the parameter
\begin{eqnarray}
\mu=\frac{m\mathcal{E}}{2Mb}+\frac{l}{2}.
\label{5.6}
\end{eqnarray}

The solution to Eq. (\ref{5.5}) is given by considering first the wave function to be regular at the origin; thus, it can be written in the form:
\begin{eqnarray}
R\left(\xi\right)=\xi^{\frac{\left|l\right|}{2}}\,e^{-\frac{\xi}{2}}\,F\left(\xi\right).
\label{5.7}
\end{eqnarray}
Therefore, substituting (\ref{5.7}) into (\ref{5.5}), we obtain
\begin{eqnarray}
\xi\,F''+\left[\left|l\right|+1-\xi\right]\,F'+\left[\mu-\frac{\left|l\right|}{2}-\frac{1}{2}\right]\,F=0.
\label{5.8}
\end{eqnarray}

The second order differential equation (\ref{5.8}) is the confluent hypergeometric equation \cite{abra}, where $F\left(\xi\right)=\,_{1}F_{1}\left(\frac{\left|l\right|}{2}+\frac{1}{2}-\mu,\,\left|l\right|+1,\,\xi\right)$ is the confluent hypergeometric function. Note that the asymptotic behaviour of the confluent hypergeometric function is given when $\xi\rightarrow\infty$, then, we can write \cite{abra}
\begin{eqnarray}
\,_{1}F_{1}\left(\bar{A},\,\bar{B},\,\xi\right)\approx\frac{\Gamma\left(\bar{B}\right)}{\Gamma\left(\bar{A}\right)}\,e^{\xi}\,\xi^{\bar{A}-\bar{B}}\,\left(1+\mathcal{O}\left(\left|\xi\right|^{-1}\right)\right),
\label{eq:}
\end{eqnarray}
where $\Gamma\left(\bar{A}\right)$ is the Gamma function. Therefore, the confluent hypergeometric function becomes well-behaved at $\xi\rightarrow\infty$ by imposing that $\bar{A}=\frac{\left|l\right|}{2}+\frac{1}{2}-\mu=-\bar{n}$ ($\bar{n}=0,1,2,\ldots$). Hence, by using Eq. (\ref{5.6}), we have
\begin{eqnarray}
\mathcal{E}_{\bar{n},\,l}=\frac{2\,M\,b}{m}\left[\bar{n}+\frac{\left|l\right|}{2}-\frac{l}{2}+\frac{1}{2}\right],
\label{5.9}
\end{eqnarray}
where the analogue of the cyclotron frequency is 
\begin{eqnarray}
\omega=\frac{2\,M\,b}{m}.
\label{5.10}
\end{eqnarray}

Despite the presence of a time-dependent magnetic field breaks the conditions for achieving the Landau quantization for a moving particle possessing a magnetic quadrupole moment, we have shown a particular case where the time-dependent magnetic field does not interact with the magnetic quadrupole moment and a discrete set of energy levels is obtained. Note that the energy levels given in Eq. (\ref{5.9}) have an infinity degeneracy as in the analogue of the Landau levels obtained in Ref. \cite{fb2,fb3}.

\section{hard-wall confining potential}

Let us consider a case of an atom with a magnetic quadrupole moment in the presence of a time-dependent magnetic field subject to a hard-wall confining potential. It is worth pointing out that in condensed matter physics, hard-wall confining potentials are used with the purpose of describing a more realistic geometry of quantum dots and quantum rings as shown in Refs. \cite{mag,mag2,dot,dot2,fur,fur2}. Therefore, let us assume that the wave function of the atom is well-behaved at the origin, and thus vanishes at a fixed radius $\xi_{0}$, i.e.,  
\begin{eqnarray}
R\left(\xi_{0}\right)=0,
\label{2.1}
\end{eqnarray}
where $\xi_{0}=M\,b\,\rho^{2}_{0}\,<\,\infty$. As we have seen previously, the parameter $\xi$ is defined in a two-dimensional system in the range $0\,<\,\xi\,<\,\infty$. On the other hand, when the system is subject to a hard-wall confining potential, then, the parameter $\xi$ becomes defined in the range $0\,<\,\xi\,<\,\xi_{0}$. By following Ref. \cite{abra}, for fixed $\xi_{0}$ and $\bar{B}$ and large $\bar{A}$, the confluent hypergeometric function can be written in the form \cite{abra}:     
\begin{eqnarray}
\,_{1}F_{1}\left(\bar{A},\,\bar{B},\,\xi_{0}\right)\approx\frac{\Gamma\left(\bar{B}\right)}{\sqrt{\pi}}\,e^{\xi_{0}/2}\,\left[\frac{\bar{B}\xi_{0}}{2}-\bar{A}\xi_{0}\right]^{\frac{1-\bar{B}}{2}}\,\cos\left(\sqrt{2\bar{B}\xi_{0}-4\bar{A}\xi_{0}}-\frac{\bar{B}\pi}{2}+\frac{\pi}{4}\right),
\label{2.2}
\end{eqnarray}
where $\Gamma\left(\bar{B}\right)$ is the Gamma function. By substituting Eqs. (\ref{2.2}) and (\ref{5.7}) into Eq. (\ref{2.1}), we obtain
\begin{eqnarray}
\mathcal{E}_{\bar{n},\,l}\approx\frac{1}{2m\rho_{0}^{2}}\left[\bar{n}\pi+\left|l\right|\frac{\pi}{2}+\frac{3\pi}{4}\right]^{2}-\frac{1}{2}\,\omega\,l.
\label{2.3}
\end{eqnarray}

The spectrum of energy (\ref{2.3}) corresponds to the energy levels of bound states for an atom with a magnetic quadrupole moment in the presence of a time-dependent magnetic field subject to a hard-wall confining potential. As we have seen in the previous section, the magnetic quadrupole moment of the neutral particle (\ref{5.1a}) does not interact with the time-dependent magnetic field, but it does with the uniform effective magnetic field (\ref{5.1}). Despite the presence of a uniform effective magnetic field characteristic of the Landau quantization, the influence of the hard-wall confining potential changes the spectrum of energy, where the energy levels (\ref{2.3}) are proportional to $\bar{n}^{2}$ (the quantum number associated with radial modes) in contrast to the analogue of the Landau levels given in Eq. (\ref{5.9}) in which is proportional to $\bar{n}$.

\section{scalar potential proportional to the inverse of the radial distance}

In this section, our focus is on the effects of a static scalar potential proportional to the inverse of the radial distance on the atom with a magnetic quadrupole moment in the presence of a time-dependent magnetic field. For this purpose, let us write the scalar potential as
\begin{eqnarray}
V\left(\rho\right)=\frac{\alpha}{\rho}
\label{6.1}
\end{eqnarray} 
where $\alpha$ is a constant that characterizes the scalar potential proportional to the inverse of the radial distance. It is worth pointing out that in the context of condensed matter physics, scalar potentials proportional to the inverse of the radial distance have been studied with 1-dimensional systems \cite{ct4,ct8,ct9,ct10,ct11}, molecules \cite{molecule,ct5,ct6}, pseudo-harmonic interactions \cite{ct12,ct13}, position-dependent mass systems \cite{pdm2,pdm3,pdm5}, the Kratzer potential \cite{kratzer,kratzer2,kratzer3}. Other contexts are the propagation of gravitational waves \cite{ct14}, quark models \cite{quark}, atoms with magnetic quadrupole moment \cite{fb} and relativistic quantum mechanics \cite{ct15,vercin,mhv,eug,cab3}.

Hence, by following the steps from Eq. (\ref{1.1}) to Eq. (\ref{5.4}), then, the radial equation (\ref{5.4}) in the presence of the scalar potential (\ref{6.1}) becomes
\begin{eqnarray}
\left[2m\mathcal{E}+2\,M\,b\,l\right]R=-R''-\frac{1}{\rho}\,R'+\frac{l^{2}}{\rho^{2}}\,R+M^{2}\,b^{2}\,\rho^{2}\,R+\frac{2m\alpha}{\rho}\,R,
\label{6.2}
\end{eqnarray}
where we have also taken $k=0$. Now, we perform a change of variable given by $r=\sqrt{M\,b}\,\,\rho$, and then, rewrite Eq. (\ref{6.2}) as
\begin{eqnarray}
R''+\frac{1}{r}\,R'-\frac{l^{2}}{\rho^{2}}\,R-r^{2}\,R-\frac{\nu}{r}\,R+\beta\,R=0,
\label{6.3}
\end{eqnarray}
where 
\begin{eqnarray}
\nu=\frac{2\,m\,\alpha}{\sqrt{M\,b}};\,\,\,\,\,\,\beta=\frac{1}{M\,b}\left[2m\mathcal{E}+2\,M\,b\,l\right].
\label{6.4}
\end{eqnarray}

Observe that the asymptotic behaviour is determined for $r\rightarrow0$ and $r\rightarrow\infty$, then, we can write the function $R\left(r\right)$ in terms of an unknown function $G\left(r\right)$ as follows \cite{mhv,vercin,heun,fb}:
\begin{eqnarray}
R\left(r\right)=e^{-\frac{r^{2}}{2}}\,r^{\left|l\right|}\,G\left(r\right).
\label{6.5}
\end{eqnarray}

By substituting the function (\ref{6.5}) into Eq. (\ref{6.3}), we obtain the following equation for $G\left(r\right)$:
\begin{eqnarray}
G''+\left[\frac{2\left|l\right|+1}{r}-2r\right]\,G'+\left[\beta-2-2\left|l\right|-\frac{\nu}{r}\right]G=0,
 \label{6.6}
\end{eqnarray}
which is called as the biconfluent Heun equation \cite{heun}, and the function $G\left(r\right)$ is the biconfluent Heun function \cite{heun}: $G\left(r\right)=H_{B}\left(2\left|l\right|,\,0,\,\beta,\,2\nu,\,r\right)$. We proceed with writing the solution to Eq. (\ref{6.6}) as a power series expansion around the origin: $G\left(r\right)=\sum_{k=0}^{\infty}a_{k}\,r^{k}$ \cite{arf,grif}. By substituting this series into Eq. (\ref{6.6}), we obtain the recurrence relation:
\begin{eqnarray}
a_{k+2}=\frac{\nu}{\left(k+2\right)\left(k+2+2\left|l\right|\right)}\,a_{k+1}-\frac{\left(\beta-2-2\left|l\right|-2k\right)}{\left(k+2\right)\left(k+2+2\left|l\right|\right)}\,a_{k},
\label{6.7}
\end{eqnarray}
and $a_{1}=\frac{\nu}{\left(1+2\left|l\right|\right)}\,a_{0}$.

By focusing on achieving bound states solutions, then, we need to impose that the biconfluent Heun series becomes a polynomial of degree $n$. From the recurrence relation (\ref{6.7}), we have that the biconfluent Heun series becomes a polynomial of degree $n$ by imposing that \cite{eug,fb}:
\begin{eqnarray}
\beta-2-2\left|l\right|=2n;\,\,\,\,\,a_{n+1}=0,
\label{6.9}
\end{eqnarray}
where $n=1,2,3,\ldots$. With the first condition given in Eq. (\ref{6.9}), we obtain 
\begin{eqnarray}
\mathcal{E}_{n,\,l}=\varpi\left[n+\left|l\right|-l+1\right],
\label{6.10}
\end{eqnarray}
where $n=1,2,3,\ldots$ is the quantum number associated with the radial modes, $l=0,\pm1,\pm2,\ldots$ is the angular momentum quantum number and the angular frequency of the system becomes
\begin{eqnarray}
\varpi=\frac{M\,b}{m}.
\label{6.10a}
\end{eqnarray}

On the other hand, in order to analyse the condition $a_{n+1}=0$, we need to obtain some coefficients of the power series expansion. We start with $a_{0}=1$, then, from Eq. (\ref{6.7}), we can obtain, for instance, the coefficients $a_{1}$ and $a_{2}$:
\begin{eqnarray}
a_{1}=\frac{\nu}{\left(1+2\left|l\right|\right)};\,\,\,\,\,\,a_{2}=\frac{\nu^{2}}{2\,\left(2+2\left|l\right|\right)\left(1+2\left|l\right|\right)}-\frac{\left(\beta-2-2\left|l\right|\right)}{2\left(2+2\left|l\right|\right)}.
\label{6.8}
\end{eqnarray}
Then, let us analyse the condition $a_{n+1}=0$ given in Eq. (\ref{6.9}) for the lowest energy state of the system ($n=1$). For $n=1$ we have that $a_{n+1}=a_{2}=0$, and thus, we obtain the relation: 
\begin{eqnarray}
\varpi_{1,\,l}=\frac{2m\alpha^{2}}{\left(1+2\left|l\right|\right)}.
\label{6.11}
\end{eqnarray}
The relation (\ref{6.11}) is obtained by assuming that the angular frequency $\varpi$ given in Eq. (\ref{6.10a}) is a parameter that can be adjusted (since we can adjust the intensity of the induced electric field (\ref{5.2})), and then, the condition $a_{n+1}=0$ is satisfied. As a result of this assumption, we have that both conditions imposed in Eq. (\ref{6.9}) are satisfied and a polynomial solution to the function $G\left(r\right)$ is obtained. Moreover, Eq. (\ref{6.11}) shows us that only some specific values of the angular frequency $\varpi$ are allowed in the system in order that bound state solutions can be obtained. These possible values of the angular frequency $\varpi$ are determined by the quantum numbers $\left\{n,\,l\right\}$ of the system and the parameter associated with the scalar potential (\ref{6.1}). For this reason, we have labelled $\varpi=\varpi_{n,\,l}$ in Eq. (\ref{6.11}) and from now on. By substituting (\ref{6.11}) into Eq. (\ref{6.10}), we have that the energy of the lowest energy state of the system is given by
\begin{eqnarray}
\mathcal{E}_{1,\,l}=\frac{2m\alpha^{2}}{\left(1+2\left|l\right|\right)}\,\left[\left|l\right|-l+2\right].
\label{6.12}
\end{eqnarray}

Hence, the general expression for the energy levels for an atom with a magnetic quadrupole moment in the presence of a time-dependent magnetic field under the influence of a scalar potential proportional to the inverse of the radial distance can be written as
\begin{eqnarray}
\mathcal{E}_{n,\,l}=\varpi_{n,\,l}\left[n+\left|l\right|-l+1\right].
\label{6.13}
\end{eqnarray}

In contrast to Eq. (\ref{5.9}), we have that the energy levels of the atom with a magnetic quadrupole moment in the presence of a time-dependent magnetic field are modified due to the influence of the scalar potential proportional to the inverse of the radial distance. Besides, the angular frequency is also modified due to the effects of the scalar potential in contrast to the cyclotron frequency $\omega$ given in Eq. (\ref{5.10}), where the possible values of the angular frequency $\varpi$ are determined by the quantum numbers of the system $\left\{n,\,l\right\}$ and the parameter associated with the scalar potential. The lowest energy state of the system becomes determined by the quantum number $n=1$ and the degeneracy of the analogue of the Landau levels is broken.

\section{linear scalar potential}

Our focus is now on the effects of a linear scalar potential on the atom with a magnetic quadrupole moment in the presence of a time-dependent magnetic field. Therefore, we introduce the following scalar potential into the Schr\"odinger equation (\ref{5.3}):
\begin{eqnarray}
V\left(\rho\right)=\eta\,\rho,
\label{7.1}
\end{eqnarray} 
where $\eta$ is a constant that characterizes the linear scalar potential. It is worth emphasizing that it has attracted a great deal of study in molecular and atomic physics \cite{linear3a,linear3b,linear3c,linear3d,linear3e,linear3f} and systems with a position-dependent mass \cite{bah}. Hence, by following again the steps from Eq. (\ref{1.1}) to Eq. (\ref{5.4}), then, we obtain the radial equation (\ref{5.4}), 
\begin{eqnarray}
\left[2m\mathcal{E}+2\,M\,b\,l\right]R=-R''-\frac{1}{\rho}\,R'+\frac{l^{2}}{\rho^{2}}\,R+M^{2}\,b^{2}\,\rho^{2}\,R+2m\,\eta\,\rho\,R.
\label{7.2}
\end{eqnarray}
where we have taken $k=0$ again. Let us perform the same change of variable made in the previous section, i.e., $r=\sqrt{M\,b}\,\,\rho$, and then
\begin{eqnarray}
R''+\frac{1}{r}\,R'-\frac{l^{2}}{\rho^{2}}\,R-r^{2}\,R-\theta\,r\,R+\beta\,R=0,
\label{7.3}
\end{eqnarray}
where the parameter $\beta$ has been defined in Eq. (\ref{6.4}) and the parameter $\theta$ is defined as
\begin{eqnarray}
\theta=\frac{2\,m\,\eta}{\left(M\,b\right)^{3/2}}.
\label{7.4}
\end{eqnarray}

By analysing the asymptotic behaviour as in the previous section, we can write the function $R\left(r\right)$ in terms of an unknown function $H\left(r\right)$ as follows \cite{mhv,vercin,heun,fb}:
\begin{eqnarray}
R\left(r\right)=e^{-\frac{r^{2}}{2}}\,e^{-\frac{\theta\,r}{2}}\,r^{\left|l\right|}\,H\left(r\right).
\label{7.5}
\end{eqnarray}

By substituting the function (\ref{7.5}) into Eq. (\ref{7.3}), we obtain the following equation for $H\left(r\right)$:
\begin{eqnarray}
H''+\left[\frac{2\left|l\right|+1}{r}-\theta-2r\right]\,H'+\left[\beta+\frac{\theta^{2}}{4}-2-2\left|l\right|-\frac{\theta\left(2\left|l\right|+1\right)}{2r}\right]H=0,
 \label{7.6}
\end{eqnarray}
which is called as the biconfluent Heun equation \cite{heun}, and the function $H\left(r\right)$ is the biconfluent Heun function \cite{heun}: $H\left(r\right)=H_{B}\left(2\left|l\right|,\,\theta,\,\beta+\frac{\theta^{2}}{4},\,0,\,r\right)$. Further, by following the steps from Eq. (\ref{6.7}) to Eq. (\ref{6.8}), we obtain a recurrence relation given by
\begin{eqnarray}
a_{k+2}=\frac{\theta\left(2k+3+2\left|l\right|\right)}{2\left(k+2\right)\left(k+2+2\left|l\right|\right)}\,a_{k+1}-\frac{\left(4\beta+\theta^{2}-8-8\left|l\right|-8k\right)}{4\left(k+2\right)\left(k+2+2\left|l\right|\right)}\,a_{k},
\label{7.7}
\end{eqnarray}
$a_{1}=\frac{\theta}{2}\,a_{0}$ and the biconfluent Heun series becomes a polynomial of degree $n$ when:
\begin{eqnarray}
4\beta+\theta^{2}-8-8\left|l\right|=8n;\,\,\,\,\,a_{n+1}=0,
\label{7.9}
\end{eqnarray}
where $n=1,2,3,\ldots$. The first condition given in Eq. (\ref{7.9}) yields 
\begin{eqnarray}
\mathcal{E}_{n,\,l}=\varpi\left[n+\left|l\right|-l+1\right]-\frac{\eta^{2}}{2m\varpi^{2}},
\label{7.10}
\end{eqnarray}
where $n=1,2,3,\ldots$ is also the quantum number associated with the radial modes, $l=0,\pm1,\pm2,\ldots$ is the angular momentum quantum number and $\varpi$ is the angular frequency of the system given in Eq. (\ref{6.10a}).

As discussed in the previous section, in order that the condition $a_{n+1}=0$ can be analysed, let us assume that $a_{0}=1$, and then, we obtain from Eq. (\ref{7.7}) a new set of coefficients of the biconfluent Heun series:
\begin{eqnarray}
a_{1}=\frac{\theta}{2};\,\,\,\,\,\,a_{2}=\frac{\theta^{2}\left(2\left|l\right|+3\right)}{8\,\left(2+2\left|l\right|\right)}-\frac{\left(4\beta+\theta^{2}-8-8\left|l\right|\right)}{8\left(2+2\left|l\right|\right)}.
\label{7.8}
\end{eqnarray}
By choosing the lowest energy state of the system, the condition $a_{n+1}=0$ yields:  
\begin{eqnarray}
\varpi_{1,\,l}=\left[\frac{\eta^{2}}{2m}\left(2\left|l\right|+3\right)\right]^{1/3},
\label{7.11}
\end{eqnarray}
where we also have that the possible values of the angular frequency are determined by the quantum numbers of the system $\left\{n,\,l\right\}$ and, in this case, by the parameter associated with the linear scalar potential. Besides, from Eqs. (\ref{7.10}) and (\ref{7.11}), the energy level associated with the lowest energy state is 
\begin{eqnarray}
\mathcal{E}_{1,\,l}=\left[\frac{\eta^{2}}{2m}\left(2\left|l\right|+3\right)\right]^{1/3}\times\left[\left|l\right|-l+2\right]-\frac{\eta^{2}}{2m}\left(\frac{2m}{\eta^{2}\left[2\left|l\right|+3\right]}\right)^{2/3}.
\label{7.12}
\end{eqnarray}

Hence, the general form the energy levels of the atom with a magnetic quadrupole moment in the presence of a time-dependent magnetic field under the influence of a linear scalar potential can be written as 
\begin{eqnarray}
\mathcal{E}_{n,\,l,\,s}=\varpi_{n,\,l}\left[n+\left|l\right|-l+1\right]-\frac{\eta^{2}}{2m\,\varpi_{n,\,l}^{2}}.
\label{7.13}
\end{eqnarray}

By comparing with the analogue of the Landau levels (\ref{5.9}), we have that the energy levels are modified by the linear scalar potential. The effects of the linear scalar potential yield a new contribution to the energy levels given by the last term of Eqs. (\ref{7.10}) and (\ref{7.13}). The angular frequency is also modified due to the linear potential in contrast to the cyclotron frequency $\omega$ given in Eq. (\ref{5.10}), but it has the same form of that obtained in Eq. (\ref{6.10a}). Furthermore, in this case, the possible values of the angular frequency $\varpi$ are determined by the quantum numbers of the system $\left\{n,\,l\right\}$ and by the parameter associated with the linear potential. Finally, the degeneracy of the analogue of the Landau levels is also broken.

\section{conclusions}

We have investigated the behaviour of an atom with a magnetic quadrupole moment in a region with a time-dependent magnetic field. We have shown that an analogue of the Landau quantization can be achieved due to the interaction of the magnetic quadrupole moment of the atom with the induced electric field. In this particular case, we have seen that a time-independent Schr\"odinger equation can be obtained, i.e., without existing the interaction between the magnetic quadrupole moment of the atom and the time-dependent magnetic field, then, bound states solutions can be obtained. 

Furthermore, we have analysed the influence of a hard-wall confining potential, a scalar potential proportional to the inverse of the radial distance and a linear scalar potential. In the case of the hard-wall confining potential, we have seen that the spectrum of energy is modified in contrast to the Landau-type levels, where the energy levels are parabolic with respect to the quantum number associated with the radial modes. In the cases of scalar potential proportional to the inverse of the radial distance potential and the linear scalar potential, we have obtained two different spectrum of energies. However, in both cases, the lowest energy state of the system becomes determined by the quantum number $n=1$ instead of the quantum number $\bar{n}=0$ obtained in the Landau-type levels, and the degeneracy of the analogue of the Landau levels is broken. Besides, the angular frequency of the system is modified in the two cases by the influence of the scalar potentials, where the possible values of the angular frequency are determined by the quantum numbers $\left\{n,\,l\right\}$ of the system and by the parameters associated with the scalar potentials. Despite the electric field given in Eq. (\ref{5.2}) to be very hard to achieve with the present technology, this work brings a new discussion about neutral particle systems that possess multipole moments.

\acknowledgments

The authors would like to thank the Brazilian agencies CNPq and CAPES for financial support.

\end{document}